\documentclass[aps,pra,reprint,superscriptaddress]{revtex4-2}
\usepackage{blindtext}
\usepackage{amsmath}
\usepackage{amssymb}
\usepackage{bbm}

\usepackage{setspace}
\usepackage{xcolor}
\usepackage{transparent}
\usepackage{comment}
\usepackage{graphicx}

\usepackage{todonotes}

\usepackage{pgf}
\usepackage{layouts}

\usepackage[font=small,justification=Justified]{caption}

\makeatletter
\newcommand{\ShowFontSize}{\f@size pt}
\makeatother

\everymath{\medmuskip=1.5mu minus 1.5mu\thickmuskip=2mu minus 2mu}

\graphicspath{{images/}}

\begin{document}
\title{Non-Hermitian topology and skin modes in the continuum via parametric processes}
\author{Markus Bestler}
\affiliation{Department of Physics, University of Konstanz, 78464 Konstanz, Germany.}
\author{Alexander Dikopoltsev}
\affiliation{Institute for Quantum Electronics, ETH Zurich, CH-8093 Z\"urich, Switzerland.}
\author{Oded Zilberberg}
\affiliation{Department of Physics, University of Konstanz, 78464 Konstanz, Germany.}

\begin{abstract}
    We demonstrate that Hermitian, nonlocal parametric pairing processes can induce non-Hermitian topology and skin modes, offering a simple alternative to complex bath engineering. Our model, stabilized by local dissipation and operating in the continuum limit, reveals exceptional points that spawn a tilted diabolical line in the dispersion. Local dissipation prevents instabilities, while a bulk anomaly signals unscreened current response. Upon opening the boundaries, we observe a non-Hermitian skin effect with localized edge modes. Through bulk winding indices and non-Bloch theory, we establish a robust bulk-boundary correspondence, highlighting parametric drives as a scalable route to non-Hermitian topology in bosonic systems.
\end{abstract}
\maketitle

Topological phases of matter constitute a remarkable class of systems whose physical behavior is governed not by local order parameters but by global, quantized invariants~\cite{TKNN_Thouless1982,resta2007theory,xiao2010berry,hasan2010colloquium,qi2011topological,bernevig2013topological,ozawa2019topological}. These topological indices characterize the bulk band structure and dictate quantized response functions, such as conductance, which are therefore immune to disorder and other local perturbations. Crucially, the robustness of these responses is often tied to the presence of anomalies, i.e., violation of conservation laws~\cite{frohlich2023gauge}. At interfaces between topologically distinct media, such as between a nontrivial phase and the vacuum, these anomalies manifest in robust boundary phenomena including gapless edge modes or fractionalized excitations. Prominent examples include the quantum Hall effect~\cite{HallEffect_Klitzing1980,TKNN_Thouless1982}, where the Hall conductance is quantized by a Chern number;  topological insulators, protected by time-reversal symmetry~\cite{hasan2010colloquium,qi2011topological}; high-order topological insulators associated to topology from higher dimensions than that of the system~\cite{resta2007theory,teo2010topological,kraus2013four,benalcazar2017quantized,petrides2018six,petrides2020higher}; and even topological phases in quasiperiodic systems~\cite{kraus2012topological,FibonacciPump_Zilberberg2015,zilberberg2021topology}.

In recent years, the study of non-Hermitian (NH) topological systems has uncovered a wealth of phenomena in open and dissipative settings, where energy is not conserved and systems are described via effective NH Hamiltonians~\cite{Masahito2020NHphysics, Kunst2021EPs, Bagarello2022NonHermitianMasterEq, FoaTorres2018NHedgestates}. In these systems, conventional topology must be reconsidered: spectra are generally complex, eigenstates lose orthogonality, and boundary sensitivity increases significantly. A striking example is the non-Hermitian skin effect (NHSE), where many bulk eigenstates localize at system edges, defying Hermitian bulk-boundary expectations~\cite{Nelson1996Hatano-NelsonModel, Zhong2018NHEdgeStates, Bergholtz2018BiorthogonalBBC, Thomale2019SkinModes}. This signals a breakdown of standard topological principles and calls for refined theoretical approaches, e.g., adaptation of topological winding numbers to the complex energy plane~\cite{SpectralWindingNumber,Sato2019SymmetryAndTopology, Fang2020WindingNumberAndCurrent}. Similarly, non-Bloch band theory further addresses the breakdown of Bloch’s theorem by introducing a generalized Brillouin zone with complex momenta, thereby restoring a modified bulk-boundary correspondence~\cite{Yokomizo2019NonBloch, Yokomizo2021BdGNonBloch, Wang2024NonBlochContinuum}. Together, these tools allow a systematic understanding of NH topology and edge behavior.

Non-Hermitian topology has been realized across a variety of experimental platforms, typically by coupling lattice systems to engineered environments. A key feature in many of these systems is the presence of exceptional points—spectral degeneracies where eigenvalues and eigenvectors coalesce—which can lead to instabilities unless mitigated by finite mode lifetimes. Consequently, most realizations rely on complex bath engineering and are modeled using tight-binding lattices. In cold atom setups, loss is induced through coupling to untrapped states or near Feshbach resonances~\cite{Williams1999FeshbachResonance, Gong2020NHSEWithColdAtomLoss}. Photonic systems employ gain and absorption via pumping and loss mechanisms, enabling topological lasing~\cite{Segev2018TopoLasingTheo, Satoshi2020ActiveTopoPhotonics, Khajavikhan2018TopoLasingExp, Kante2017NonreciprocalTopoLasing}, directional amplification and non-reciprocal transport~\cite{Valle2015LightTransport, Metelmann2015Nonreciprocal, Nunnenkamp2020Amplification, Szameit2020FunnelingOfLight}. Electronic circuits use resistors and amplifiers to implement asymmetric couplings~\cite{Thomale2024CircuitRealization}, while mechanical and acoustic platforms achieve similar effects through damping and motor-induced drives~\cite{Coulais2020NHMetaMaterial}. Still, the need for tailored baths remains a major obstacle, emphasizing the need for more natural and scalable implementations of non-Hermitian topology.

\begin{figure}[!t]
\includegraphics[width=1\columnwidth]{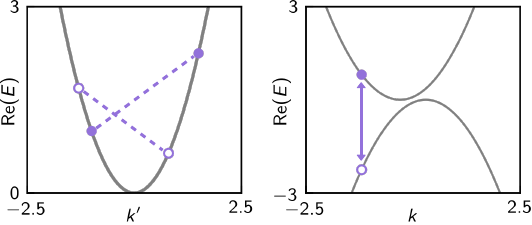}
\caption{\textit{Model}. (a) Sketch of the model [cf.~Eq.~\eqref{eq:MainHamiltonian}]: bosonic particles with a parabolic dispersion (grey line) are created  (annihilated) in pairs with opposite momentum plus a momentum shift $2k_0$, e.g., marked by full (empty) purple circles connected by dashed lines. This also leads to an energy shift between created (annihilated) particle pairs. (b) The dynamical matrix of the system over a shifted momentum Nambu spinor [cf.~Eq.~\eqref{eq:DynamicalMatrix}]. Here, the pair creation (annihilation) appear as a particle-hole scattering with a  shifted momentum. We use $k_0=0.3$.}
\label{Fig1}
\end{figure}

Parametric driving offers a powerful route to engineer effective interactions and dissipation in quantum systems~\cite{eichler2023classical}. Typically realized through two-photon (or two-mode) processes, such drives can induce squeezing, implement coherent pairing terms, or mimic effective damping, all while preserving Hermitian dynamics. By tuning the drive amplitude and frequency, these systems can reach exceptional points, beyond which time-translation symmetry breaking bifurcations emerge. Such instabilities are often stabilized and postponed by local dissipation~\cite{Soriente2021DissipationInducedPT}. These features make parametric drives especially valuable in systems where direct coupling or dissipation control is challenging. Prominent applications include parametric amplifiers (paramps) for quantum-limited signal readout~\cite{Siddiqi2015parametricAmplifier}, optomechanical sensors with enhanced sensitivity~\cite{2015ParametricLIGO}, and Kerr parametric oscillators used in quantum information processing~\cite{Grimm_2019Kerr-cat},  and as Ising annealers~\cite{HeugelIsing2022, Ameye2025InstabilityLandscape, DykmanInteraction2018}. 

In this work, we demonstrate that Hermitian, nonlocal parametric pairing processes can give rise to non-Hermitian topology and skin modes, without relying on complicated bath engineering. Our model is stabilized by simple local dissipation and operates in the continuum, free from any underlying lattice structure. We begin by analyzing the bulk of the continuum system, where the parametric drive leads to level attraction between particle and hole branches. Beyond a critical threshold, exceptional points emerge and form a diabolical line in the complex spectrum. Crucially, local dissipation stabilizes the system, preventing dynamical instabilities. We then identify a bulk anomaly that signals an unscreened current response, hinting at anomalous boundary behavior. Indeed, upon opening the boundaries, we observe a clear non-Hermitian skin effect with strongly localized edge modes. We quantify this behavior by constructing a bulk winding index and complete the topological characterization using non-Bloch band theory, which restores a consistent bulk-boundary correspondence for our continuum-driven model. These findings establish parametric drives as a minimal and scalable route to realizing non-Hermitian topology in bosonic systems.

We consider a continuum one-dimensional bosonic model with momentum-shifted particle pair production [see Fig.~\ref{Fig1}(a)]
\begin{align}
    \resizebox{0.9\hsize}{!}{$\displaystyle{H=\int dk' \left[\frac{k'^2}{2m} a_{k'}^\dagger a_{k'}^{\phantom \dagger} + \left(ig \,a_{k'}^\dagger a_{-k'-2k_0}^{ \dagger} + h.c.\right) \right]\, .}$}
    \label{eq:MainHamiltonian}
\end{align}
The particles have a  parabolic dispersion as a function of wavenumber $k'$, and we use units where $\hbar=1$. For simplicity, we set $m=1/4$. The bosonic annihilation operator $a_{k'}^{\phantom \dagger}$ removes a particle from mode $k'$. Pairs of particles are created (annihilated) in a purely imaginary parametric process with amplitude $ig$, involving opposing momenta alongside a momentum shift of $2k_0$, akin to finite-momentum cooper pairs~\cite{FiniteMomentumCooperPairs}. It is comfortable, henceforth, to shift the momentum origin to $k=k'+k_0$. Thus, we can rewrite Eq.~\eqref{eq:MainHamiltonian} on top of a shifted momentum bosonic Nambu spinor $\mathbf{a}_k = (\hat{a}_{k-k_0}, \hat{a}_{-k-k_0}^\dagger)^T$ with an associated Hamiltonian $H=\int dk\, \mathbf{a}_k^\dagger \mathbf{H}_k \mathbf{a}_k$, where $\mathbf{H}_k$ is a $2\times 2$ Hamiltonian density.  In order to diagonalize the model, we write the corresponding dynamical matrix~\cite{DynamicalMatrix}
\begin{align}
    \mathbf{D}_k\equiv\sigma_z \mathbf{H}_k=
    \begin{pmatrix}
        (k-k_0)^2 & ig\\
        ig & -(-k-k_0)^2
    \end{pmatrix}
    \, ,
    \label{eq:DynamicalMatrix}
\end{align}
where $\sigma_z$ is a Pauli matrix. Here, the pair creation (annihilation) maps to a particle-hole scattering process, see Fig.~\ref{Fig1}(b). 

\begin{figure*}[!ht]
\includegraphics[width=1\textwidth]{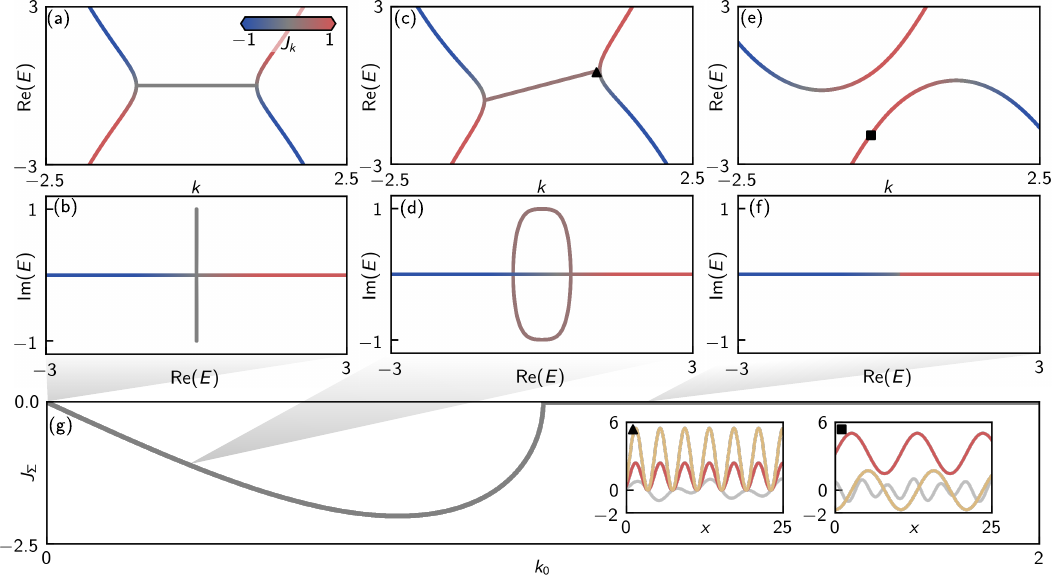}
\caption{\textit{Bulk spectrum and anomaly}. (a), (c), and (e) Dispersion relation, $\rm{Re}(E)$, as a function of the shifted momentum $k$, for $\gamma=g=1$ and $k_0=0,0.3, 1.2$, respectively. Modes are colored according to the probability current $J_k$ [ cf.~Eq.~\eqref{eq:ContinuityEquation}]. (b), (d), and (f) Spectral plot of lifetime $\mathrm{Im}(E)$ as a function of $\mathrm{Re}(E)$ for the same parameters and coloring as in (a), (c) and (e), respectively. (g) Total current of modes not participating in the diabolical line, $J_\Sigma$, as a function of the momentum shifts $k_0$. Insets: Wavefunction amplitude $|\psi_k(x)|$ (grey), probability current $J$ (red), and interaction-generated sink/source term (yellow) in real space [cf.~Eq.~\eqref{eq:ContinuityEquation}] corresponding to the modes marked with a triangle in (c) and square in (d).}
\label{Fig2}
\end{figure*}

The resulting spectrum reads $E=-2k k_0 \pm \sqrt{(k^2+k_0^2)^2-g^2}$, see Fig.~\ref{Fig2} (a)-(f). When $g=0$, we have free parabolic particle and hole dispersions with an indirect gap, due to the shifted momentum labeling. As the particle-hole scattering is purely imaginary, it leads to level attraction~\cite{LevelAttraction}. Hence, we observe three distinct regimes:
(i) for $k_0=0$ and $g\neq0$, no momentum shift is present in the system. However, due to the level attraction, the particle and hole bands coalesce for $|k|<\sqrt{|g|}$, forming a horizontal diabolical line terminated by two exceptional points, see Fig.~\ref{Fig1}(a). For the modes on the diabolical line, $\mathrm{Im}(E)$ splits into two branches with a lifetime of $\mathrm{Im}(E) \lessgtr 0$, marking  parametric instabilities,  see Fig.~\ref{Fig2}(b). (ii) for finite $|k_0|<\sqrt{|g|}$, the diabolical line is still present for $|k|<\sqrt{|g|-k_0^2}$, see Fig.~\ref{Fig1}(c). Crucially, it now becomes tilted due to the momentum shift. Additionally, in the spectrum of the system, the two branches with nonzero $\mathrm{Im}(E)$ form two closed loops, see Fig.~\ref{Fig1}(d). (iii) increasing $k_0$ further until $|k_0|>\sqrt{|g|}$, the momentum shift is sufficiently strong  compared with the particle-hole coupling to open the indirect gap as in the non-interacting case, see Fig~\ref{Fig2}(e). The diabolical line is not present anymore and the free dispersion is now only marginally affected. Note that, as the diabolical line vanishes, all eigenergies are purely real, see Fig~\ref{Fig2}(f). 
As the level attraction in the system can lead to the appearance of a diabolical line and therefore negative lifetimes [$\mathrm{Im}(E)>0)$], see Fig.~\ref{Fig2}(b) and (d), we introduce a uniform local damping of strength $\gamma$ in the system. The Hamiltonian of the system then becomes non-Hermitian and reads $\tilde{H}=H-2i\gamma\int dk'a_{k'}^\dagger a_{k'}^{\phantom\dagger}$ with a corresponding dynamical matrix $\tilde{\mathbf{D}}_k=\sigma_z\mathbf{H}_k-i\gamma\mathbbm{1}$. In a physical system, we require the damping to be sufficiently strong, $\gamma\geq\sqrt{g^2-k_0^4}$, such that all the modes would have positive lifetimes.
%

As the scattering particles and holes are shifted in momentum, the diabolical line is tilted, i.e., $d[\mathrm{Re}(E)]/dk\neq 0$ for $k_0\neq0$ . In a Hermitian system, this would imply that these modes have a finite group velocity, i.e., these modes contribute to a finite probability current. To study whether this behavior extends to the non-Hermitian case, we write the effective von Neumann equation for our non-Hermitian model $\tilde{H}$ (see Appendix \ref{AppendixA})
\begin{align}
    \partial_t\rho  = -2\partial_x J-2\gamma\rho+4g\mathrm{Re}(e^{-2ik_0x}\psi^2)\,,
    \label{eq:ContinuityEquation}
\end{align}
for the density $\rho=\psi^\dagger(x)\psi(x)$, where $\psi(x)=(1/\sqrt{2\pi})\int dk e^{ikx}a_k$ is the local bosonic field operator, and  
the closed system probability current reads $J=-i[(\partial_x \psi)\psi^\dagger-(\partial_x \psi^\dagger)\psi]$. Compared with the standard Hermitian continuity equation, Eq.~\eqref{eq:ContinuityEquation} features additional sink and source terms. An obvious sink arises due to the uniform damping,  $2\gamma\rho$. The term arising due to the parametric pair production, $\propto\mathrm{Re}(e^{2ik_0x}\psi^2)$, acts as a spatially-dependent sink or source, depending on the spatial structure of the wavefunction $\psi_k(x)$ of a mode $k$, see insets of Fig.~\ref{Fig2}.

The contribution from the local sinks and sources, arising from the pairing term, averages out for modes that are not part of the diabolical line, namely, when integrating over distances in space much larger than the periodicity of the wave function.
Hence, for long length scales, these modes experience the same density loss. It is instructive to define $J_\Sigma$ as the total current arising from these modes, see Fig.~\ref{Fig2}(g). 
In the case of weak coupling, $\sqrt{|g|}<|k_0|$, we have no diabolical line modes, and $J_\Sigma=\int_{-\infty}^\infty J_k dk$ sums over the current $J_k$ arising from all modes of the system. In this limit, every mode in the system has a counter-propagating mode [see Fig~\ref{Fig2}(e)], regardless of the choice of the global damping $\gamma$, such that $J_\Sigma=0$. Instead, when the diabolical line appears, $\sqrt{|g|}>|k_0|$, we have $J_\Sigma=\int_{-\infty}^{-\sqrt{|g|-k_0^2}} J_k dk + \int_{\sqrt{|g|-k_0^2}}^{\infty} J_k dk\neq0$, which may be understood as an anomaly that must be screened by the modes in the diabolical line. Indeed, as long as the diabolical line is tilted, its modes transport a finite probability current, see Fig.~\ref{Fig2}(a) and (c). However, as $\mathrm{Re}(e^{-2ik_0x}\psi^2)$ does not average out over space for these modes, they experience additional sink or source terms, see insets of Fig.~\ref{Fig2}(g). Their transported current therefore changes relative to the other modes'  currents over time. As a result, modes on the diabolical line cannot act as counter-propagating modes that screen the nonvanishing $J_\Sigma\neq 0$. Thus, we obtain a true anomaly in the bulk, which requires a corresponding boundary effect that will cancel the unscreened current in the bulk~\cite{frohlich2023gauge}. Such bulk-boundary correspondence implies that when open boundary conditions are applied, boundary modes must appear in the system.

\begin{figure*}[!ht]
\includegraphics[width=1\textwidth]{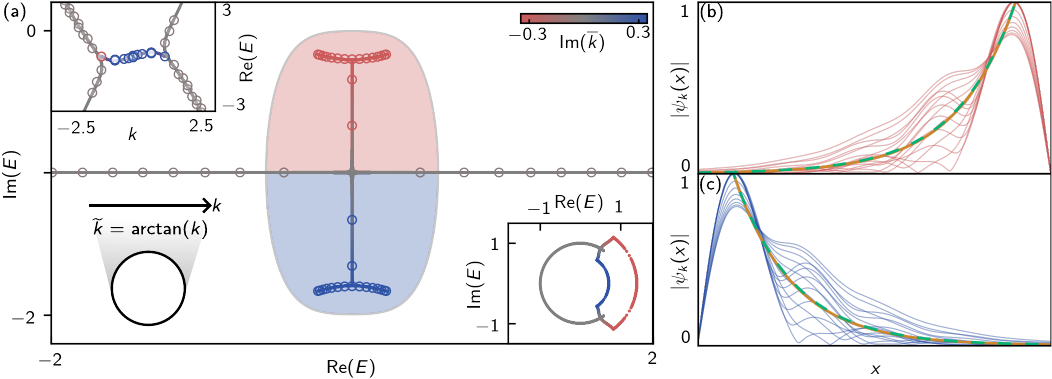}
\caption{\textit{Bulk-boundary correspondence}. Spectral plot as in Fig.~\ref{Fig1}(d) for $k_0=0.3$ and $g=\gamma=1$ calculated numerically for a system with open boundary conditions (empty circle markers) and via non-Bloch theory (full markers). Modes localized on the left (right) edge are colored red (blue). Delocalized modes are colored grey. Colors of the non-Bloch theory calculation correspond to $\mathrm{Im}(\bar k)$. Areas marked in light red (light blue) correspond to energies with a spectral winding number of $+1$ ($-1$) [cf.~Eq.~\eqref{eq:WindingNumber}]. Energies outside of these two areas have spectral winding number $0$. Insets (counter clockwise, starting from top left): dispersion relation calculated numerically [cf.~Fig.~\ref{Fig1}(c)] (empty circle markers) and via non-Bloch theory (full markers); Illustration of how the continuous bulk momentum, $k$, can be compactified to a periodic momentum $\tilde k$ by mapping it to a circle via $\tilde k=\arctan(k)$; and generalized Brillouin zone of the model in terms of the non-Bloch factor $\beta$. (b) Local density of states of numerically calculated finite-frequency skin modes on the right edge of the open system (red) and exponential $e^{\mathrm{Im}(\bar{k}_l)(x-x_0)}$ with the localization $\mathrm{Im}(k_l)$ of the most (orange, $\mathrm{Im}(\bar{k_l})\approx0.345$) and the least (green dashed, $\mathrm{Im}(\bar{k_l})\approx0.339$) localized finite frequency skin modes. (c) Same as in (b) for the left edge of the system.}
\label{Fig3}
\end{figure*}
%

To confirm the formation of boundary modes, we Fourier transform the dynamical matrix $\tilde{\mathrm{D}}_k$ to real space, then discretize it and  diagonalize the finite system numerically, see the eigenvalues of such a system in the upper left inset in Fig.~\ref{Fig3}(a). We ascribe a wavenumber to each mode by Fourier transforming it and picking the highest amplitude Fourier component as $k$. The resulting dispersion relation has a similar structure as the bulk continuous model, cf.~Fig.~\ref{Fig1}(c). Unlike the bulk model, however, the complex energy spectrum at the diabolical line forms two arches with $\mathrm{Im}(E)\lessgtr-\gamma$ instead of closed loops, cf.~Fig.~\ref{Fig3}(a) and   Fig.~\ref{Fig2}(c), respectively. Moreover, the modes with $\mathrm{Im}(E)>-\gamma$ [$\mathrm{Im}(E)<-\gamma$] are exponentially localized at the right [left] boundary of the system for $k_0>0$, see e.g., numerically calculated modes in  Figs.~\ref{Fig3}(b) and (c).
  
The accumulation of a large number of modes near the boundaries and the strong sensitivity of the energy spectrum to boundary conditions is known as the non-Hermitian skin effect (NHSE). Crucially, in this work, the NHSE arises from the Hermitian coupling between momentum-shifted particle and hole modes, which gives rise to the tilted diabolical line; the non-Hermitian damping only ensures a positive lifetime for all modes.
The NHSE has a topological origin that can be classified via a spectral winding number~\cite{SpectralWindingNumber} 
\begin{align}
W(E) = \int_{-\pi}^{\pi} \frac{d \tilde{k}}{2\pi i}\frac{d}{d \tilde{k}}\log(\det(\tilde{\mathbf{D}}_{\tilde{k}}(\tilde{k})-E))\,,
\label{eq:WindingNumber}
\end{align}
for periodic tight-biding systems with a Brillouin zone $\tilde{k}\in[-\pi,\pi]$, where $E\in\mathbb{C}$ is a complex energy, and $W(E)>0$ ($W(E)<0$) implies boundary modes on the right (left) side of the system. In our continuum model~\eqref{eq:MainHamiltonian}, $k\in(-\infty, \infty)$. To apply the spectral winding number, we compactify momentum space using $\tilde{k}=\arctan(k)$, and wrap $-\infty$ onto $\infty$ to obtain a periodic structure~\cite{CompactificationChernNumber}, see Fig.~\ref{Fig3}(a), bottom left. This approach guarantees a well quantized topological index, as the system approaches the thermodynamic limit, $k\to\pm\infty$.  
Calculating the winding number for every $E\in\mathbb{C}$ reveals that the diabolical line engenders two regions with $W(E)=\pm 1$ separated by a line at $\mathrm{Im}(E)=-\gamma$, see Fig.~\ref{Fig3}(a). We thus find the topological index classifying the  bulk anomaly presented in Fig.~\ref{Fig2}(g).

The spectral Winding number~\eqref{eq:WindingNumber} predicts localized skin modes in a semi-infinite system. In Fig.~\ref{Fig3}(a), we see that the system with open boundary conditions on both ends of the chain exhibits skin modes.
To prove the non-Hermitian bulk-boundary correspondence in the finite model, we use non-Bloch theory~\cite{Yokomizo2019NonBloch}. This approach also allows us to analytically describe the localization lengths of the skin modes. To apply non-Bloch theory, we replace the real wavenumber $k$ in Eq.~\eqref{eq:DynamicalMatrix} with a complex wavenumber $\bar{k}\in\mathbb{C}$. Here, $\mathrm{Im}(\bar k)$ describes the spatial localization of a mode. To select the modes which satisfy the open boundary conditions, we first solve the non-Bloch equation
\begin{align}
    \mathrm{det}(\tilde{\mathbf{D}}_{\bar{k}}-\mathbbm{1} E)=0
    \label{eq:NonBlochEquation}
\end{align}
for every $E\in\mathbb{C}$. We obtain a fourth order polynomial equation in $\bar{k}$, which has four solutions $\bar{k}_j$ with $j\in\{1,2,3,4\}$ for every $E$. We sort the solutions such that $\mathrm{Im}(\bar{k}_j)\leq\mathrm{Im}(\bar{k}_{j+1})$. The non-Bloch condition states that the modes $\bar{k}_2$ and $\bar{k}_3$ fulfill the open boundary conditions, if and only if $\mathrm{Im}(\bar{k}_2)=\mathrm{Im}(\bar{k}_3)$. For the modes $\bar{k}_1$ and $\bar{k}_4$, $\mathrm{Im}(\bar{k}_1)=\mathrm{Im}(\bar{k}_2)=\mathrm{Im}(\bar{k}_3)=\mathrm{Im}(\bar{k}_4)$ have to hold to fulfill the open boundary conditions. The collection of all $\bar{k}$ satisfying this requirement constitute the generalized Brillouin zone. Commonly, the generalized Brillouin zone is visually represented in terms of the non-Bloch factor $\beta=\exp(i \bar{k}z)$, see bottom right inset of Fig.~\ref{Fig3}(a). In our case, it takes the form of a unit circle, missing one section at the diabolical line. At the missing section, the generalized Brillouin zone splits up into two arcs corresponding to the modes localized on the different edges of the system. Using Eq.~\eqref{eq:NonBlochEquation},  we also find the open boundary energy spectrum of the model, see Fig~\ref{Fig3}(a). The exact spectrum coincides with the numerical one. In the top left inset of Fig.~\ref{Fig3}(a), we draw the obtained dispersion relation, which is in agreement with the numerical solutions.

Having successfully applied non-Bloch theory to our model, we harness it to find the explicit localization, $|\mathrm{Im}(\bar k)|$, of the skin modes analytically, see Figs.~\ref{Fig3}(b) and (c). In our case, the non-Bloch equation~\eqref{eq:NonBlochEquation} is a fourth order polynomial, which is analytically solvable. However, the eigenmodes also have to obey the non-Bloch condition, rendering this a challenging task to solve analytically. We therefore focus on calculating the localization of finite-frequency modes with highest and lowest $|\mathrm{Im}(\bar k)|$. For details on this calculation and how to also find the localization of zero-frequency modes, see Appendix~\ref{AppendixB}.
The most localized modes [highest $|\mathrm{Im}(\bar{k})|$] with finite frequency  are those with $\mathrm{Re}(E)\to 0$. In this limit, the non-Bloch equation~\eqref{eq:NonBlochEquation} and the non-Bloch condition can be combined and reduce the problem to a third-order polynomial in $\mathrm{Im}(\bar k)^2$. Solving for its roots leaves us with an analytical expression for $\mathrm{Im}(\bar k)$. As an example, for $g=\gamma=1$ and $k_0=0.3$ , we obtain $|\mathrm{Im}(\bar{k})|\approx0.345$, see Figs.~\ref{Fig3}(b) and (c).
Inspecting the solutions of Eq.~\eqref{eq:NonBlochEquation}, we find that for $|k_0/\sqrt{|g|}|\lesssim 0.35$, the least localized modes [lowest $|\mathrm{Im}(\bar{k})|$] with finite frequency fulfill the non-Bloch condition by a double zero. Therefore $\bar k_2=\bar k_3$, instead of only $\mathrm{Im}(\bar k_2)=\mathrm{Im}(\bar k_3)$ holds. Hence, we can use the discriminant of Eq.~\eqref{eq:NonBlochEquation} and combine the non-Bloch equation with the non-Bloch condition to obtain an analytically solvable quartic equation. For $g=\gamma=1$ and $k_0=0.3$, this yields a localization of $|\mathrm{Im}(\bar{k})|\approx0.339$.
We find the same localization strengths on both edges. The explicit calculation of the localization properties of the skin modes in a finite system, therefore completes the proof of the non-Hermitian bulk-boundary correspondence for our system.

We introduced a continuum model that exhibits topological skin modes arising from nonlocal parametric processes. This mechanism, fundamentally distinct from conventional non-Hermitian models, enables the realization of directional localization in systems lacking lattice periodicity. Our framework is broadly applicable to a variety of bosonic platforms, including ultracold atoms, polaritonic condensates, and acoustic metamaterials. Moreover, it provides a natural route to realizing non-Hermitian topology in the Bogoliubov excitations of solitonic backgrounds, both in static and time-dependent settings. Future work will explore how strong interactions modify and enrich the topological structure of the model, potentially leading to new regimes of nonlinear topological dynamics.

\section*{Acknowledgements}
We are grateful for fruitful discussions with J.~del Pino, B.~Schneider and J.~Faist. We acknowledge funding from the Deutsche Forschungsgemeinschaft (DFG) via project numbers 449653034; 425217212; 521530974; 545605411;  and through SFB1432, as well as from the Swiss National Science Foundation (SNSF) through the Sinergia Grant No.~CRSII5\_206008/1.ETH Fellowship program: (22-1 FEL-46) (to AD).

\bibliography{bibliography.bib}

\clearpage

\appendix

\section{Derivation of the non-Hermitian continuity equation}
\label{AppendixA}

The effective von Neumann equation corresponding to Eq.~\eqref{eq:MainHamiltonian} in the main text is $\dot \rho = -i[\int dk \mathbf{a}_\mathbf{k}^\dagger \mathbf{H}_\mathbf{k}^{\phantom\dagger} \mathbf{a}_\mathbf{k}^{\phantom \dagger}, \rho]-2i\gamma \rho$. Here, the first term engenders the Hermitian dynamics of the system, while the second term describes local damping. Note, that such a damping term can be derived by taking the mean-field limit of a Lindblad master equation with single-particle jump operators~\cite{Bagarello2022NonHermitianMasterEq}. To derive Eq.~\eqref{eq:ContinuityEquation} in the main text, we use the definitions $\rho=\psi^\dagger(x)\psi(x)$ and $\psi(x)=(1/\sqrt{2\pi})\int dk e^{ikx} a_k$ to transform the effective von-Neumann equation to real space. To separate the effects of parabolic dispersion and momentum-shifted coupling, we split the model into $H=\int dk \mathbf{a}_\mathbf{k}^\dagger \mathbf{H}_\mathbf{k}^{\phantom\dagger}\mathbf{a}_\mathbf{k}^{\phantom \dagger}= H_{disp}+H_{coupl}$.
\begin{widetext}
The parabolic dispersion $H_{disp}=\int_{-\infty}^\infty (k-k_0)^2 a_{k-k_0} a^\dagger_{k-k_0} +(-k-k_0)^2 a^\dagger_{-k-k_0} a_{-k-k_0} dk$ leads to the closed system probability current: 
\begin{align}
\dot \rho =& -i[H_{disp}, \rho]
\nonumber
\\
=& -i[\int_{-\infty}^\infty (k-k_0)^2 a_{k-k_0} a^\dagger_{k-k_0} +(-k-k_0)^2 a^\dagger_{-k-k_0} a_{-k-k_0}dk,\int_{-\infty}^\infty e^{-ik'x}a_{k'}^\dagger dk'\int_{-\infty}^\infty e^{ik''x}a_{k''} dk'' ]
\nonumber
\\
=&
-i\int_{-\infty}^\infty\int_{-\infty}^\infty\int_{-\infty}^\infty [2k^2 a_{k} a^\dagger_{k}, e^{-ik'x}a_{k'}^\dagger e^{ik''x}a_{k''}]dkdk'dk''
\nonumber
\\
=&-i\int_{-\infty}^\infty\int_{-\infty}^\infty\int_{-\infty}^\infty (-2k^2 a_{k}^\dagger e^{-ik'x} \delta (k-k')e^{ik''x}a_{k''} + e^{-ik'x}a_{k'}^\dagger 2k^2 a_{k} e^{ik''x}\delta(k-k'') dkdk'dk''
\nonumber
\\
=& -2i\left(\int_{-\infty}^\infty\int_{-\infty}^\infty
(\frac{d^2}{dx^2}e^{-ikx} a_{k}^\dagger) e^{ik''x}a_{k''} dk dk'' -\int_{-\infty}^\infty\int_{-\infty}^\infty e^{-ik'x}a_{k'}^\dagger (\frac{d^2}{dx^2} a_{k} e^{ikx})
dkdk'\right)
\nonumber
\\
=& -2i \frac{d}{dx}[(\frac{d}{dx}\psi^\dagger)\psi-\psi^\dagger(\frac{d}{dx}\psi)] = -2\partial_x J\,,
\label{SIeq:ProbabilityCurrentTerm}
\end{align}
and the momentum-shifted particle-hole coupling $H_{coupl}=\int_{-\infty}^\infty ig (a_{k-k_0}a_{-k-k_0})-ig (a_{-k-k_0}^\dagger a_{k-k_0}^\dagger) dk$ generates the additional spatially-dependent sink or source term:
\begin{align}
    \dot\rho=&-i[H_{coupl}, \rho]\nonumber
    \\
    =& -i[\int_{-\infty}^\infty ig (a_{k-k_0}a_{-k-k_0})-ig (a_{-k-k_0}^\dagger a_{k-k_0}^\dagger) dk, \int_{-\infty}^\infty e^{-ik'x}a_{k'}^\dagger dk'\int_{-\infty}^\infty e^{ik''x}a_{k''} dk'']
    \nonumber
    \\
    =& g\int_{-\infty}^\infty\int_{-\infty}^\infty\int_{-\infty}^\infty [ a_{k-k_0}a_{-k-k_0}, e^{-ik'x}a_{k'}^\dagger e^{ik''x}a_{k''}]-[ a_{k-k_0}^\dagger a_{-k-k_0}^\dagger, e^{ik'x}a_{k'} e^{-ik''x}a_{k''}^\dagger]dkdk'dk''
    \nonumber
    \\
    =& g\int_{-\infty}^\infty\int_{-\infty}^\infty\int_{-\infty}^\infty (e^{-ik'x}e^{ik''x}[ a_{k-k_0}a_{-k-k_0}, a_{k'}^\dagger a_{k''}]-e^{ik'x}e^{-ik''x}[ a_{k-k_0}^\dagger a_{-k-k_0}^\dagger, a_{k'} a_{k''}^\dagger])dkdk'dk''
    \nonumber
    \\
    =& g\int_{-\infty}^\infty\int_{-\infty}^\infty\int_{-\infty}^\infty e^{-ik'x}e^{ik''x}(a_{k-k_0}\delta(-k-k_0-k')+\delta(k-k_0-k')a_{-k-k_0})a_{k''}
    \nonumber
    \\
    &+e^{ik'x}e^{-ik''x} (a_{k-k_0}^\dagger \delta(-k-k_0-k')+\delta(k-k_0-k')a_{-k-k_0}^\dagger)a_{k''}^\dagger dkdk'dk''
    \nonumber
    \\
    =& 2g(e^{2ik_0x}\psi\psi + e^{-2ik_0x}\psi^\dagger\psi^\dagger)=4g\mathrm{Re}(e^{2ik_0x}\psi\psi)\,.
    \label{SIeq:SinkSourceTerm}
\end{align}
\end{widetext}
Inserting Eqs.~\eqref{SIeq:ProbabilityCurrentTerm} and~\eqref{SIeq:SinkSourceTerm} into the effective von Neumann equation then yields Eq.~\eqref{eq:ContinuityEquation} in the main text.
\section{Details on the calculation of the localization length}
\label{AppendixB}
\subsection{Zero-frequency modes and Finite-frequency modes with strongest localization}
\begin{figure}[!ht]
\includegraphics[width=1\columnwidth]{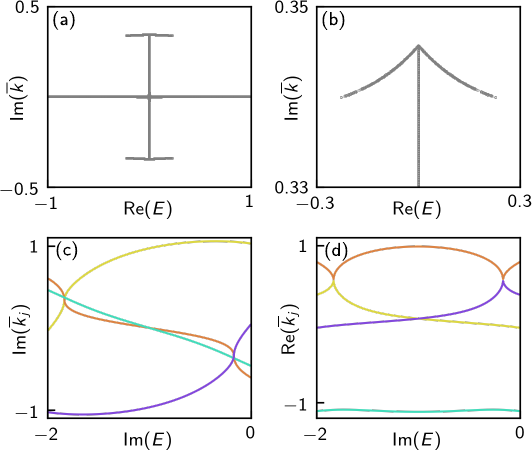}
\caption{\textit{Non-Bloch theory for $g=\gamma=1$ and $k_0=0.3$} (aFrequency $\mathrm{Re}(E)$ and localization $\mathrm{Im}(\bar k)$ of non-Bloch modes. (b) Zoom-in of (a) on the end of the top arch. (c) and (d) Imaginary and real part of all solutions $\bar{k}_j$ of Eq.~\eqref{eq:NonBlochEquation} for $\mathrm{Re}(E)=0.23$, corresponding to the finite-frequency mode with smallest $|\mathrm{Im}(\bar{k})|$ [see (b)].}
\label{FigSI1}
\end{figure}
We first focus on the zero frequency modes [$\mathrm{Re}(E)=0$]. We therefore solve the non-Bloch equation
\begin{align}
    \label{SIeq:Non-BlochEquationOnlyImag}
    \mathrm{det}(\mathbf{D}_{\bar k}-\mathrm{Im}(E))=0\,.
\end{align}
For this, we split Eq.~\ref{SIeq:Non-BlochEquationOnlyImag} into its real and imaginary parts. Solving the imaginary part for $\mathrm{Re}(\bar k)$ leads to the doubly degenrate solution $\mathrm{Re}(\bar k_1) =0$ and
\begin{align}
    \mathrm{Re}(\bar k_\pm)=-\pm\sqrt{\frac{k_0\mathrm{Im}(E)+k_0^2\mathrm{Im}(\bar k) + \mathrm{Im}(\bar k)^3+k_0\gamma}{ \mathrm{Im}(\bar k)}}\,.
    \label{SIeq:LeastLocalizedModeRealPartk}
\end{align}
\begin{widetext}
We now insert $\mathrm{Re}(\bar k)$ into the real part of Eq.~\eqref{SIeq:Non-BlochEquationOnlyImag}. For $\mathrm{Re}(\bar k_1)$, this leads to
\begin{align}
    -\gamma ^2-\mathrm{Im}(E)^2-2 \mathrm{Im}(E) (\gamma +2 k_0 \mathrm{Im}(\bar{k}_1))+g^2-\left(k_0^2+\mathrm{Im}(\bar{k}_1)^2\right)^2-4 \gamma  k_0 \mathrm{Im}(\bar{k}_1)=0
    \label{SIeq:realpartk0}\, ,
\end{align}
and for $\mathrm{Re}(\bar k_\pm)$, we get
\begin{align}
    -\left(\mathrm{Im}(E)^2 \left(k_0^2+\mathrm{Im}(\bar{k}_\pm)^2\right)\right)-2 \gamma  \mathrm{Im}(E) \left(k_0^2+\mathrm{Im}(\bar{k}_\pm)^2\right)+g^2 \mathrm{Im}(\bar{k}_\pm)^2+\left(k_0^2+\mathrm{Im}(\bar{k}_\pm)^2\right) \left(4 \mathrm{Im}(\bar{k}_\pm)^4-\gamma ^2\right) = 0\,.
    \label{SIeq:realpartkpm}
\end{align}
\end{widetext}
We find that the non-Bloch condition can be fulfilled in two different ways: (a) If $k_+$ and $k_-$ satisfy the non-Bloch condition, Eq.~\eqref{SIeq:realpartkpm} defines a curve in the $\mathrm{Im}(E)$-$\mathrm{Im}(\bar{k})$-plane. This is the curve we see in Fig.~\ref{Fig3}(a) as the straight line in the complex spectrum, since $\mathrm{Re}(E)$ is set to $0$. Or (b), the non-Bloch condition is fulfilled by one of the doubly degenerate $k_0$ and one of the $k_\pm$. This solution corresponds to the point, where modes with $\mathrm{Re}(E)=0$ modes split up into two branches on either side of the $\mathrm{Im}(E)=-\gamma$-line. Solving Eq.~\eqref{SIeq:realpartk0} for $\mathrm{Im}(E)$ therefore gives two possible solutions
\begin{align}
    \mathrm{Im}(E)=&-\gamma-2 k_0 \mathrm{Im}(\bar{k}_1)
    \nonumber
    \\
    &\pm\sqrt{g^2-k_0^4+2 k_0^2 \mathrm{Im}(\bar{k}_1)^2-\mathrm{Im}(\bar{k}_1)^4}\,.
    \label{SIeq:Im(E)HighLoc}
\end{align}
We insert these values for $\mathrm{Im}(E)$ into Eq.~\eqref{SIeq:realpartkpm} and arrive at a third order polynomial for $\mathrm{Im}(\bar{k})^2$:
\begin{align}
    -g^2 k_0^2+k_0^6+7 k_0^4 \mathrm{Im}(\bar{k})^2+31 k_0^2 \mathrm{Im}(\bar{k})^4+25 \mathrm{Im}(\bar{k})^6=0\,.
    \label{SIeq:FinalPolynomialHighLoc}
\end{align}
As a third order polynomial, the solutions of Eq.~\eqref{SIeq:FinalPolynomialHighLoc} can be given in a closed analytical formula. These modes correspond to the modes with highest $|\mathrm{Im}(\bar{k})|$ and therefore strongest localization of the finite-frequency modes, see Fig.~\ref{FigSI1}(a) and (b). Inserting $\gamma=g=1$ and $k_0=0.3$ into Eq.~\eqref{SIeq:Im(E)HighLoc} gives $\mathrm{Im}(E)\approx-1.79$ and $\mathrm{Im}(E)\approx-0.21$ respectively. Using the solutions of Eq.~\eqref{SIeq:FinalPolynomialHighLoc}, we get $\mathrm{Im}(\bar k)\approx\pm 0.345$.
\subsection{Finite-frequency modes with weakest localization}
Fig.~\ref{FigSI1}(c) and (d) show the solutions of the non-Bloch equation [Eq.~\eqref{eq:NonBlochEquation}] for the frequency of the weakest localized finite frequency mode as a function its lifetime. In the case of $|k_0/\sqrt{|g|}|\lesssim0.35$, we see that the non-Bloch condition is fulfilled by a double-zero of the non-Bloch equation. A polynomial equation having a double zero is equivalent to its discriminant equating to zero. The discriminant of the non-Bloch equation is
\begin{widetext}
\begin{align}
    -256 \tilde{E}^6-768 \tilde{E}^4 g^2-2048 \tilde{E}^4 k_0^4-768 \tilde{E}^2 g^4+5120 \tilde{E}^2 g^2 k_0^4-4096 \tilde{E}^2 k_0^8-256 g^6+256 g^4 k_0^4=0\, ,
    \label{SIeq:Discriminant}
\end{align}
\end{widetext}
where we introduced $\tilde{E}=E+i\gamma$. This discriminant then is a third order polynomial in $\tilde{E}^2$. We can therefore find a analytical expression for $\tilde{E}$. The resulting expression can be inserted in the non-Bloch equation [Eq.~\eqref{eq:NonBlochEquation}] to arrive at a fourth order polynomial for $\bar{k}$. Solving this fourth order polynomial leads to the localization of the least localized finite-frequency mode. In the case of $\gamma=g=1$ and $k_0=0.3$, using Eq.~\eqref{SIeq:Discriminant} we get $E\approx\pm0.23-1.83i$ and $E\approx\pm0.23-0.17i$. The four solutions correspond to the for ends of arches in Fig.~\ref{Fig3}. Inserting $E$ into Eq.~\eqref{eq:NonBlochEquation} gives $\mathrm{Im}(\bar{k})\approx0.339$.

\end{document}